\def\aj{{AJ}}

\def\apj{{ApJ}}
\def\apjs{{ApJS}}
\def\apjl{{ApJL}}

\def\etal{{\it et al.}}
\def\chandra{{\it Chandra\/}}
\def\XMM{{\it XMM\/}}

\documentclass[cup5b]{caps}
\usepackage{graphicx}
\usepackage{amssymb}
\usepackage{ociwsymp3e}  

\HeadText{V. E. Margoniner and the Deep Lens Survey Team} 

\def\plotone#1{\centering \leavevmode
\includegraphics[width=.95\columnwidth]{#1}}

\def\plottwo#1#2{\centering \leavevmode
\includegraphics[width=.45\columnwidth]{#1} \hfil
\includegraphics[width=.45\columnwidth]{#2}}

\def\plotthree#1#2#3{\centering \leavevmode
\includegraphics[width=.32\columnwidth]{#1} \hfil
\includegraphics[width=.32\columnwidth]{#2} \hfil
\includegraphics[width=.32\columnwidth]{#3}}

\begin{document}

\pagenumbering{arabic}

\author[]{V. E. MARGONINER$^{1}$, and the DEEP LENS SURVEY TEAM\\(1) Bell Laboratories, Lucent Technologies, NJ, USA}

\chapter{Shear-selected clusters from the \\ Deep Lens Survey}

\begin{abstract}

Weak gravitational lensing has the potential to select clusters
independently of their baryon content, dynamical state, and star
formation history.  We present steps toward the first shear-selected
sample of clusters, from the Deep Lens Survey (DLS), a deep
$BVRz^{\prime}$ imaging survey of 28 square degrees.  Cluster
redshifts are determined from photometric redshifts of members and
from lensing tomography, and in some cases have been confirmed
spectroscopically.  DLS imaging data are also used to derive
mass-to-light ratios, and upcoming Chandra and XMM time will yield
X-ray luminosities and temperatures for a subsample of 12 clusters.
Thus we can begin to address any baryon or luminous-matter bias which
may be present in current optical and X-ray samples.  When the DLS is
complete, we expect to have a sample of $\sim 100$ shear-selected
clusters from $z \sim 0.2-1$.

\end{abstract}

\section{Introduction}

Already in 1937 F. Zwicky (Zwicky 1937) found that clusters of
galaxies are mostly composed of dark matter, yet all cluster catalogs
to date have been based on detection of the luminous, baryonic,
component of clusters, which is a small fraction of the total matter.
Two recently developed methods based on cluster effects on the
background have the potential to change that.  The Sunyaev-Zel'dovich
effect (SZE) method, based on the upscattering of cosmic microwave
background photons by the intracluster medium, will select clusters
independent of their star formation history, but still depends on
baryon content and temperature (Birkinshaw 2003, Romer 2003).  Weak
gravitational lensing, in which background galaxy images are sheared
by the mass of an intervening cluster, will select clusters
independent of star formation history, baryon content, {\it and}
dynamical state.

While each technique has its strengths and weaknesses, we note that
the use of clusters as tracers of large-scale structure, and thus as
indicators of cosmological parameters such as $\Omega_m$ and
$\sigma_8$, hinges on their mass function, not merely on their number
density.  Thus it is crucial to compile samples which reflect purely
the mass function, with no dependence on star formation history or
dynamical state.  Shear-selected samples come close to that ideal.
Any difference between shear-selected and traditional samples will
also be interesting for those who study clusters as systems in their
own right.

Here we present steps toward the first shear-selected sample of
clusters, using 12 deg$^2$ of the Deep Lens Survey (DLS).  Because
many of the clusters have not yet been assigned redshifts or followed
up in other ways, in this paper we concentrate on techniques and
examples rather than aggregate features of the sample.

\section{Data}

The DLS is a deep multicolor (BVRz$^\prime$) imaging survey of 28
deg$^2$ being carried out at the 4-m telescopes of the US National
Observatories (KPNO and CTIO).  Over the course of the survey, seven
separate $2^\circ$ by $2^\circ$ fields will be imaged in B, V, R, and
z$^\prime$ to a limiting surface brightness (1$\sigma$) of
approximately 29, 29, 29, and 28 mag arcsec$^{-2}$, respectively.  The
$R$ filter is used when the seeing is 0.9$^{\prime\prime}$ or better,
to optimize its utility for lensing studies.  The other filters
provide color information for photometric redshift estimates, and are
observed when the seeing is worse than 0.9$^{\prime\prime}$.
 
The survey has been awarded approximately 100 dark nights and as of
now 75\% of the observations have been completed.  More details about
the DLS can be found in this proceedings (Margoniner \etal~2003) or
at: {\tt http://dls.bell-labs.com}.


For a uniform data set, we chose 12 deg$^2$ which had total $R$
exposure times of at least 13,500 seconds as of March 2002.  Some of
this area does not yet have $BVz'$ coverage, so only the $R$ data are
used for cluster detection in this paper.  Where $BVz'$ data are
available, they may be used to assign redshifts to detected clusters.

\section{Mass Maps}

From the stacked $R$ images (see Wittman {\it et al.} 2002 for details
on the image processing), we make preliminary catalogs using
SExtractor (Bertin \& Arnouts 1996).  These are used as input to the
{\tt ellipto} software described in Bernstein \& Jarvis (2002), which
measures weighted moments of each object.  Available subfields (40' on
a side, roughly the size of the camera) within each 2$^\circ$ field
are stitched together to make one large catalog covering anywhere from
1 deg$^2$ (for fields with only two available subfields) up to 4
deg$^2$ (for fields with all subfields available).  In the stitching
process, multiple measurements of objects in the overlap areas between
subfields are compared, and discrepant objects thrown out.  The
fraction of objects thrown out this way is small, indicating good
uniformity across subfields.


To derive a clean, high-median-redshift subcatalog for each field, we
select sources with the following properties: (1) {\tt ellipto}
errorcode of zero, (2) SExtractor flags $<4$ ({\it i.e.} split objects
allowed, but nothing with a reall error), (3) {\tt ellipto} size
$>1.2$ times the point-spread function (PSF) size, because sources
must be resolved to show a lensing signal (4) $23<R<25$, to eliminate
low-redshift galaxies without going so faint as to include noise
objects, (5) {\it ellipto} size $< 25$ pixel$^2$, also to
eliminate low-redshift (higher angular diameter) galaxies, and (6)
observed ellipticity $<0.5$, because with $\sim 1^{\prime\prime}$
resolution, highly elliptical objects are quite likely to be blends of
two distinct sources, based on our measurements of the Hubble Deep Field
and synthetic fields convolved with this seeing.

We then make two types of maps based on these catalogs.  The first is
simply a convergence ({\it i.e.}, projected mass surface density times
a quantity involving distances of sources and lenses) map based on the
algorithm of Fischer \& Tyson (1997).  The second map is a
modification of the Fischer \& Tyson algorithm, with one less power of
$r$ in the kernel, which roughly represents projected potential rather
than mass.  This second map serves as a sanity check for the first;
requiring that they appear on both types of maps reduces the false
detection rate.  In our simulations, we have found that we can detect
clusters as small as 500 km s$^{-1}$ equivalent velocity dispersion
over the redshift range 0.2--0.7 using this technique. The
highest shear clusters in our sample have equivalent velocity
dispersions of about 900 km s$^{-1}$, and they can be easily seen on
both maps.

Finally, tests of shear systematics and $RMS$ noise are made. The
first test consists of rotating by $45^\circ$ the ellipticities of all
sources. Because lensing does not affect this component of
ellipticity, the resulting ``mass'' map gives some indication of noise
and systematic errors.  The highest peaks on these maps are far below
those of the clusters.  The second test is the construction of a map
from randomized positions~({\it i.e.} the xy position of one object
becomes the position of another, which preserves any systematics which
might come from exclusion zones or the like, while nulling any real
lensing signal).  Here again, the highest peaks are much lower than on
the real mass map.

\begin{figure}
\centering
\includegraphics[width=.95\columnwidth]{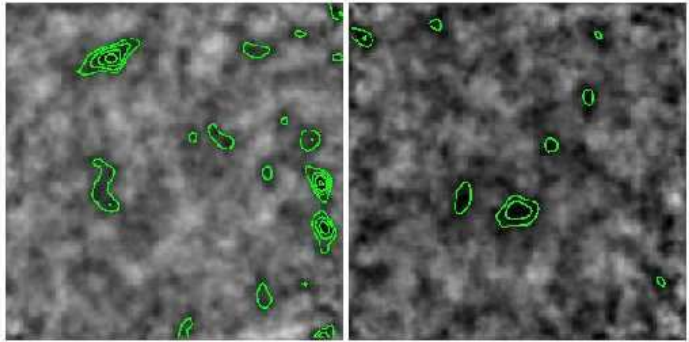} 
\caption{Mass (left) and randomized (right) maps for DLS field F3. The
contours indicate positive detections and indicate the same levels in
both maps. North is Up, East is to the Left.}
\label{fig:maps}
\end{figure}

Figure \ref{fig:maps} shows the mass map and its corresponding
randomized map for a single 4 deg$^2$ DLS field, which is not yet
imaged to full depth.  Approximately $122$K sources were used in the
construction of these $2^\circ \times 2^\circ$ maps. We expect to
double the number of sources, decreasing the noise by $\sim \sqrt{2}$,
when full-depth is achieved.  In this field, the peak at upper left is
in our Chandra sample; the other two obvious peaks would have
qualified for the sample, but were not in the 12 deg$^2$ available at
the time.  These two candidates will be in a second Chandra proposal
extending the area covered. Other peaks may be noise; judgement will
be reserved until after the field is imaged to full depth.

\section{Photometric Redshifts}

\begin{figure}
\centering \includegraphics[width=.70\columnwidth]{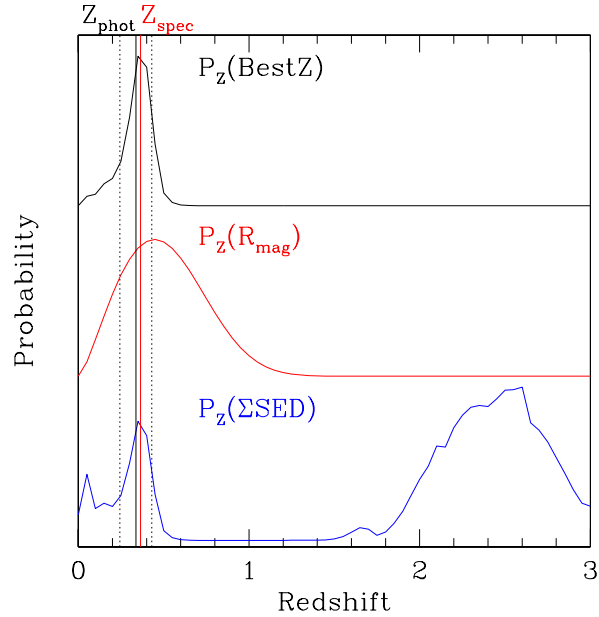}
\caption{This galaxy, classified as IE by Cohen \etal (1999) and found
to be at redshift 0.364, has $R=20.9$ and a photometric redshift
$BestZ=0.337 \pm 0.094$}
\label{fig:pmofz}
\end{figure}

The multiband ($BVRz^\prime$) DLS observations allows us to estimate
photometric redshifts for $\sim 40$ galaxies per square arc min.  Our
technique is based on SED fitting and on a luminosity function prior.
From the comparison between the observed colors of an object and the
colors expected from different galaxy types at a range of redshifts, a
probability distribution, $pc(z)$, is derived.  To compute $pc(z)$ we
use the publicly available HyperZ code (Bolzonella \etal~2000).  We
then compute the probability that an object of apparent magnitude $m$
is at redshift $z$, $pm(z)$, assuming a Schechter luminosity function
and taking into account the volume element at $z$ (Peebles 1980).  The
product of $pc(z)$ and $pm(z)$ generates a final probability
distribution from which we determine our photometric redshift (called
BestZ) and its uncertainty. Figure \ref{fig:pmofz} shows an example of
such probabilities for a galaxy in the DLS with spectroscopic redshift
and type determined by Cohen \etal (1999).


\begin{figure}
\centering
\plottwo{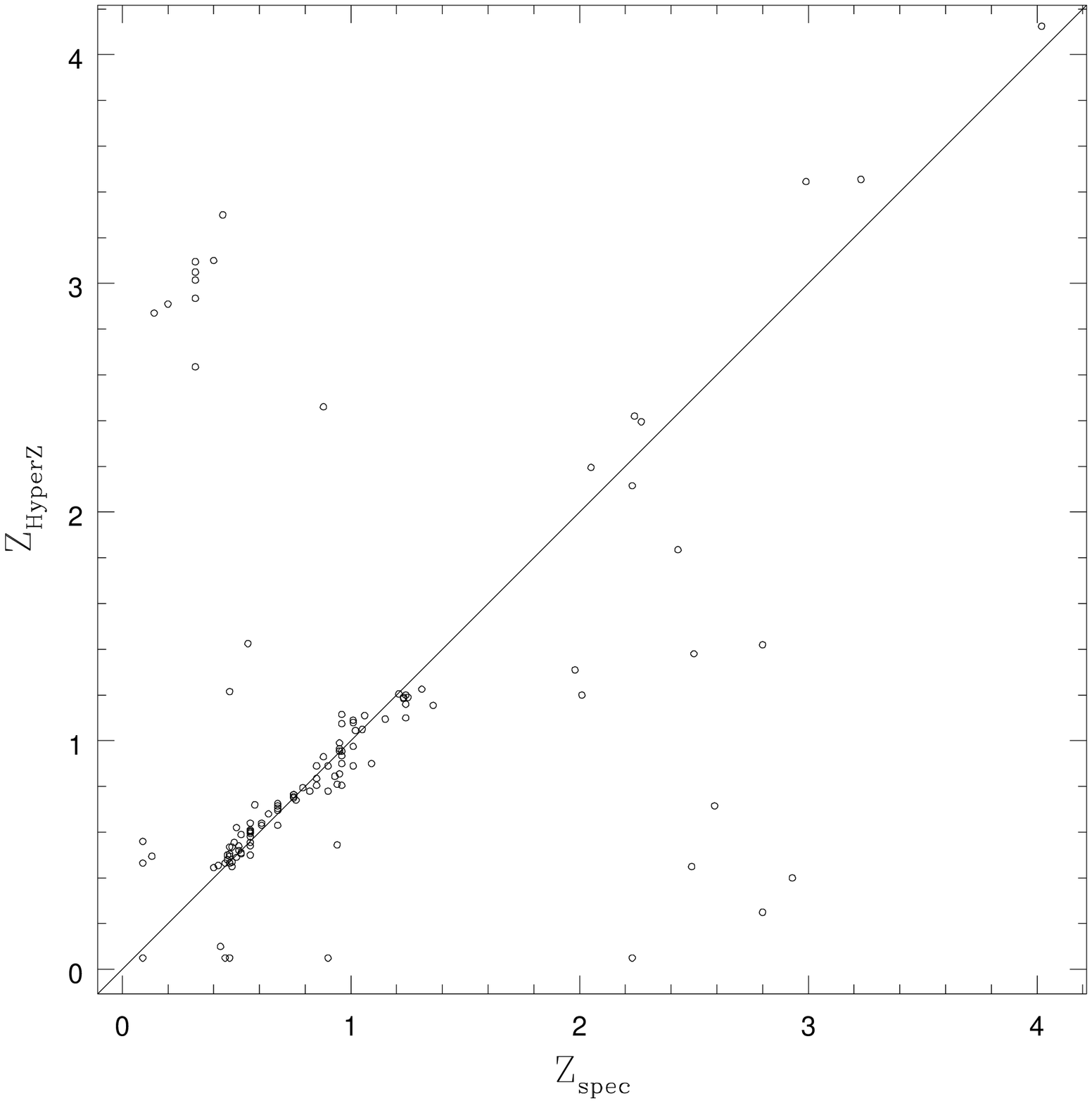}{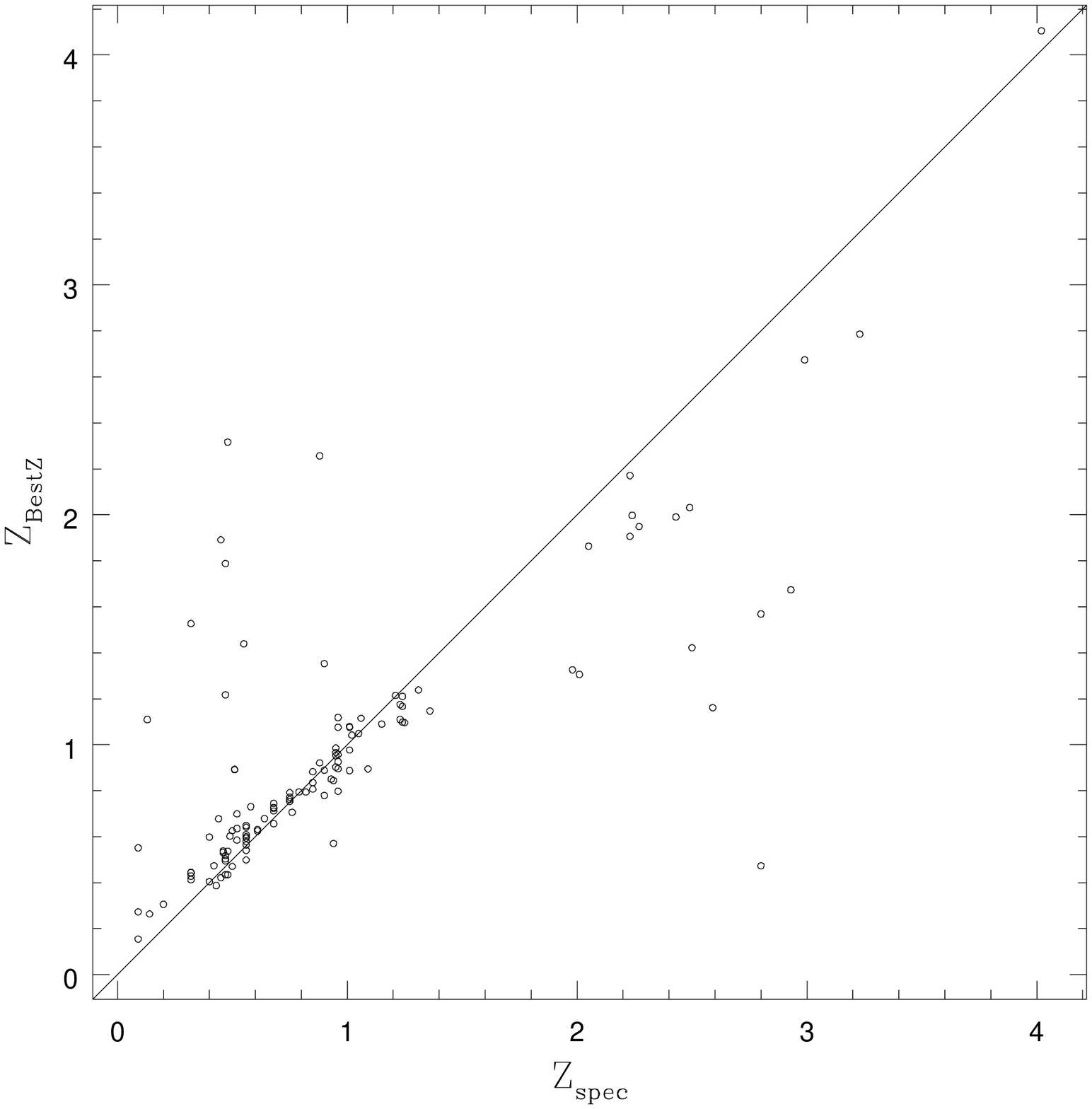}
\caption{Photometric redshifts based on 4 bands (f450w, f606w, f814w, J)  photometry of 119 galaxies in the HDFN. The left panel shows HyperZ results ($\Delta_{100\%}=-0.13\pm 0.59$, $\Delta_{95\%}=-0.03 \pm 0.37$); The right panel shows our BestZ results ($\Delta_{100\%}=-0.05 \pm 0.25$, $\Delta_{95\%}=-0.00 \pm 0.14$).}
\label{fig:hdfn4}
\end{figure}

Because of the limited number of filters in the DLS, and its
relatively small wavelength range, the inclusion of a magnitude prior
improves significantly the photometric redshifts estimates from HyperZ
alone. To illustrate this, Figure \ref{fig:hdfn4} shows spectroscopic
(from Cohen \etal~2000) versus photometric redshifts for 119 galaxies
in the HDFN, using as input photometry only a limited filter set which
roughly simulates the DLS (f450w, f606w, f814w, J).  The left panel
shows HyperZ estimates, and the right panel shows our BestZ
results. We quote our errors in terms of $1+z$: $\Delta_{100\%} =
(z_{spec}-z_{phot})/(1+z_{spec})$ takes into account all objects; and
$\Delta_{95\%}$ exclude the 5\% worst of outliers (as is often done in
the literature).  In each case, we quote a mean, which represents any
overall bias, and an rms, which is the expected error for a single
galaxy.


\begin{figure}
\centering
\plotthree{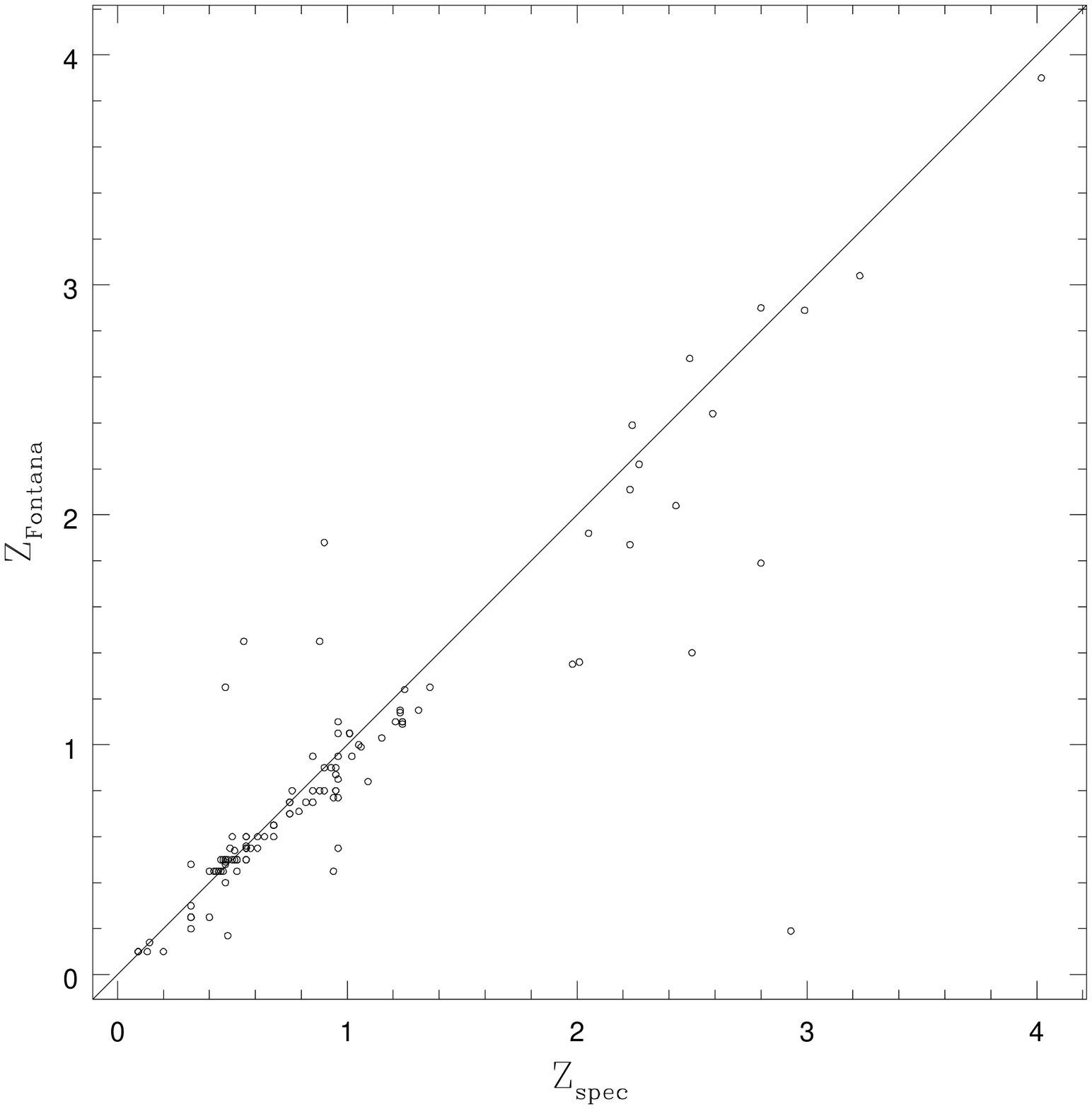}{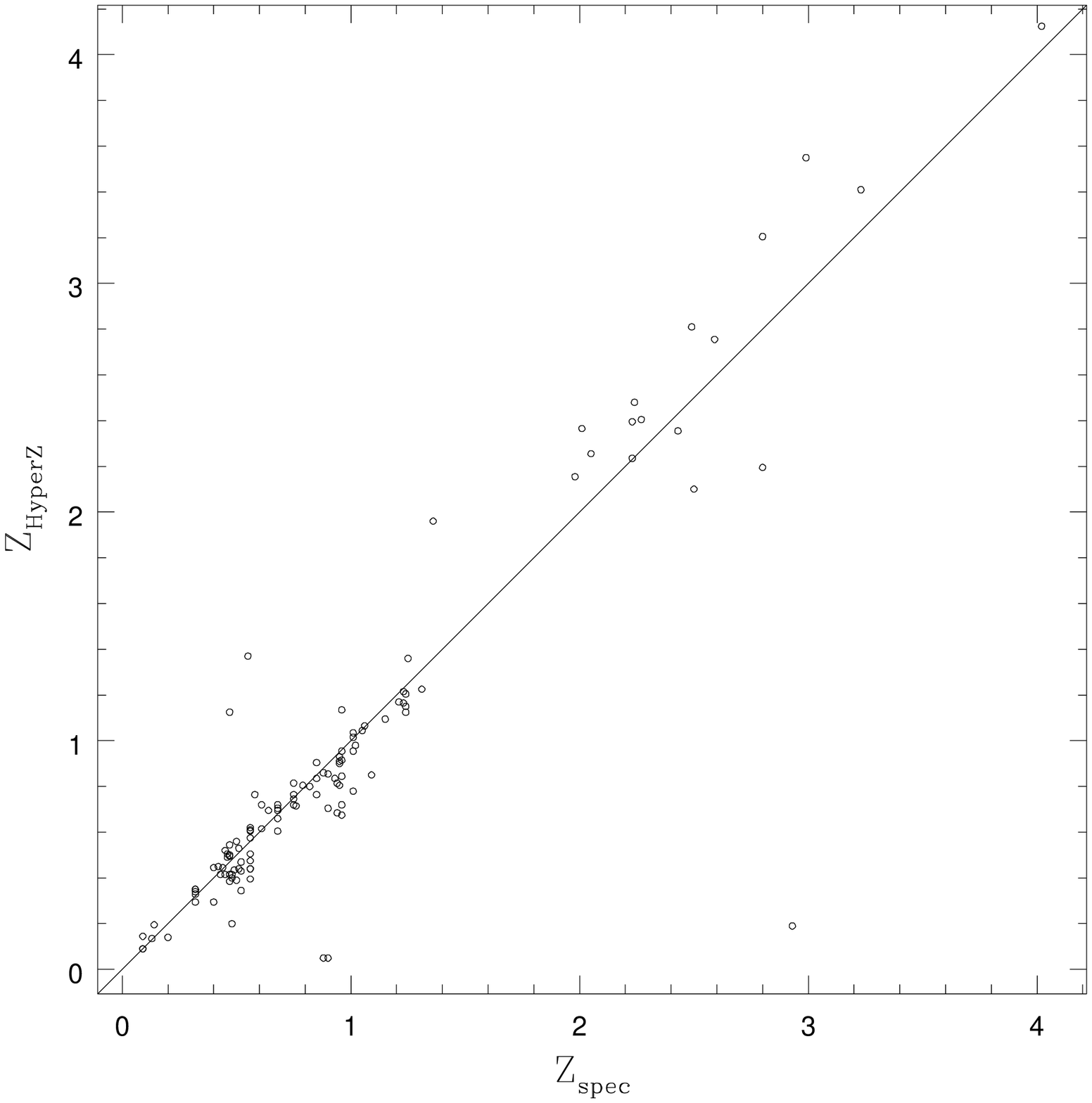}{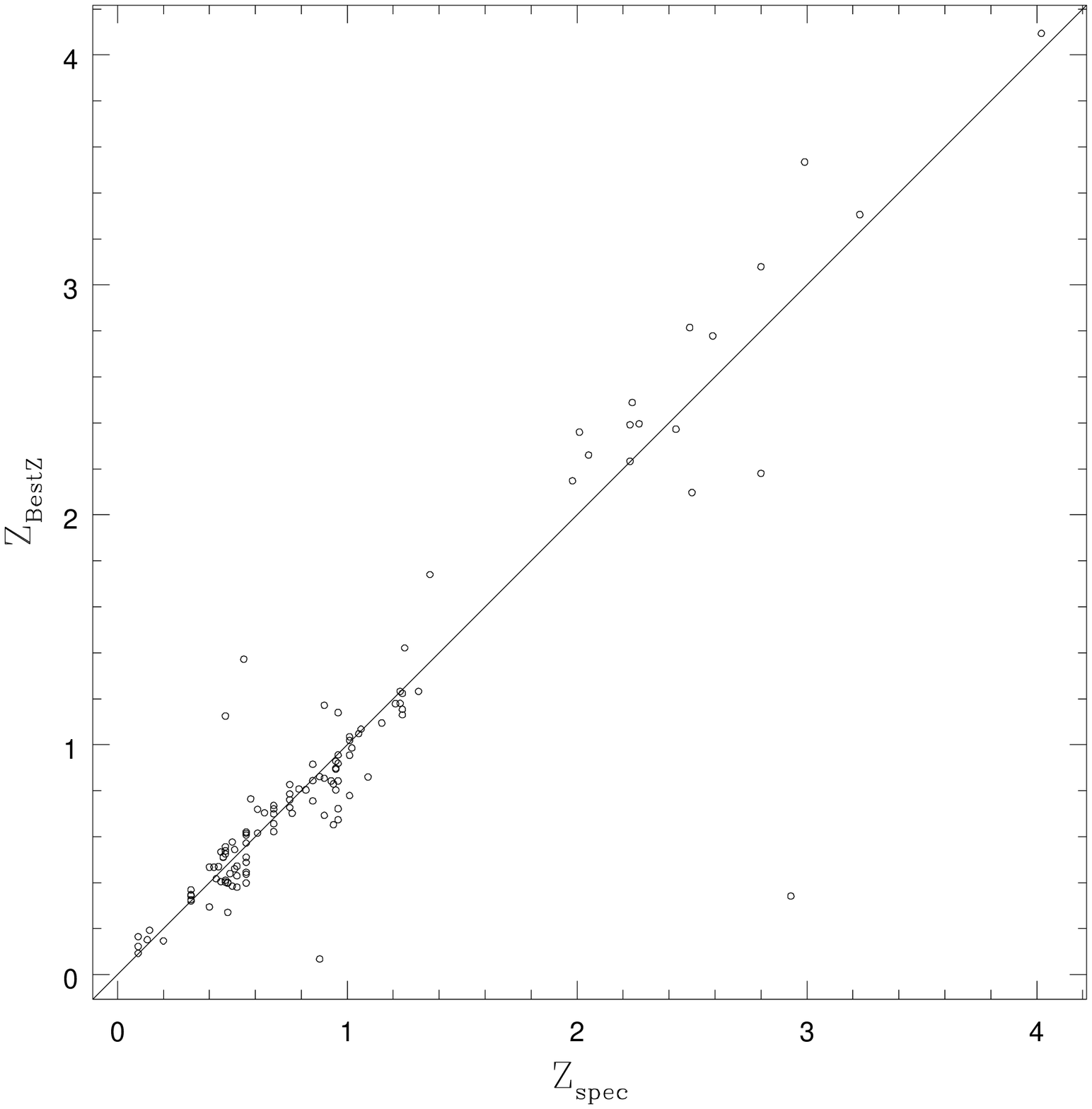}
\caption{Photometric redshifts based on 7 band photometry of 119
galaxies in the HDFN. The left panel shows Fontana \etal~ 2000
estimates ($\Delta_{100\%}=0.02 \pm 0.13$, $\Delta_{95\%}=0.03 \pm
0.06$); the middle panel shows HyperZ results ($\Delta_{100\%}=0.01 \pm
0.12$, $\Delta_{95\%}=0.01 \pm 0.06$); and the right panel shows our
BestZ results ($\Delta_{100\%}=0.00 \pm 0.11$, $\Delta_{95\%}=0.00 \pm
0.06$).}
\label{fig:hdfn7}
\end{figure}

Figure \ref{fig:hdfn7} shows photometric redshifts based on the 7-band
(f300w, f450w, f606w, f814w, J, H, K) photometry available for the
HDFN galaxies (Fontana \etal~2000). The left panel shows the often
mentioned Fontana \etal~2000 results; the middle panel shows HyperZ
estimates, and the right panel shows our BestZ estimates.  In this
case, the HyperZ estimates are nearly as good as BestZ, indicating
that the luminosity function prior is not necessary with an extensive
filter set.  At the same time, it demonstrates that our approach works
well in general; the larger per-galaxy errors in the DLS are a result
of the limited filter set, not the algorithm.  We note that the
limited DLS filter set was a conscious choice, as the per-galaxy noise
in lensing is already limited by the random orientation of each galaxy
(shape noise).

\begin{figure}
\centering 
\includegraphics[width=.45\columnwidth]{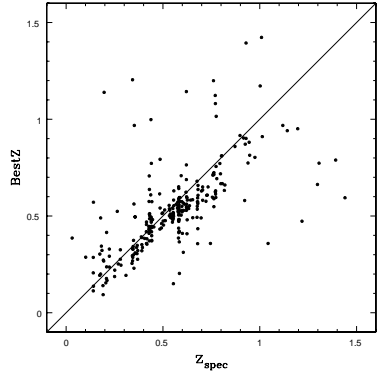}
\caption{Photometric redshifts for 271 galaxies with DLS photometry
and spectroscopic redshifts ($\Delta_{100\%}=-0,00 \pm 0.19$,
$\Delta_{95\%}=0.02 \pm 0.08$).}
\label{fig:F1p22zxz}
\end{figure}

DLS subfield F1p22 contains the Caltech Faint Galaxy Redshift Survey
(CFGRS, Cohen \etal~1999).  After excluding stars, quasars, and
anything with bad quality factor (quality>6) (see Cohen \etal~1999) we
were left with 275 galaxies with reliable redshift measurements and
DLS photometry.  Figure \ref{fig:F1p22zxz} shows our results. From
this analysis we can infer that, at least up to $z \sim 1.5$, most
photometric redshifts obtained from the DLS data will have a precison
of $\sim10\%$ in $(1+z)$.  This meets the design goal of the DLS;
shape noise, not redshift noise, dominates.

There are {\it many} parameters that affect the precision of
photometric redshifts.  The quality of the photometry, the number of
bands and its wavelength coverage, and the set of spectral energy
distribution (SED) templates are some of most important ones.  We
tested different sets of SED templates: the observed SEDs from
Coleman, Wu \& Weedman 1980; the synthetic spectra from Bruzual \&
Charlot 1993; and templates reconstructed from the colors of objects
with known redshift in the HDF and SDSS (private contribution from
Andrew Connolly).  While each set of templates produced acceptable
results, we found best results using this last set of templates,
both for HyperZ alone and for our method.

\section{Tomography}

\begin{figure}[h]
\centering \includegraphics[width=.40\columnwidth]{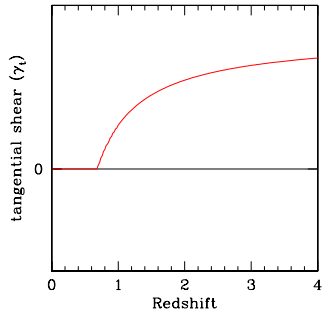}
\caption{Theoretical relation between tangential shear and source
redshift for a lens at $z=0.63$. The vertical axis is in arbitrary
units and depends on lens mass.}
\label{fig:shearxz}
\end{figure}

The dependency between the amplitude of the shear and the redshifts of
lens and source, provides an unique tool capable of estimating the
lens redshift independently of its galaxy members.  Equation 1 shows
the relation between the amplitude of the tangential shear and the
{\it angular diameter distance} from the observer to the lens
($D_{L}$), from the observer the source ($D_{S}$), and from the lens
to the source ($D_{LS}$). A lens, or galaxy cluster, at redshift
$z_{lens}$ is not capable of deforming objects at $z<z_{lens}$ and
shears more strongly objects at $z>>z_{lens}$ ($\gamma^{t} \propto
{{D_{L}D_{LS}} \over {D_{S}}}$, Figure \ref{fig:shearxz}).

\begin{figure}[h]
\centering
\plottwo{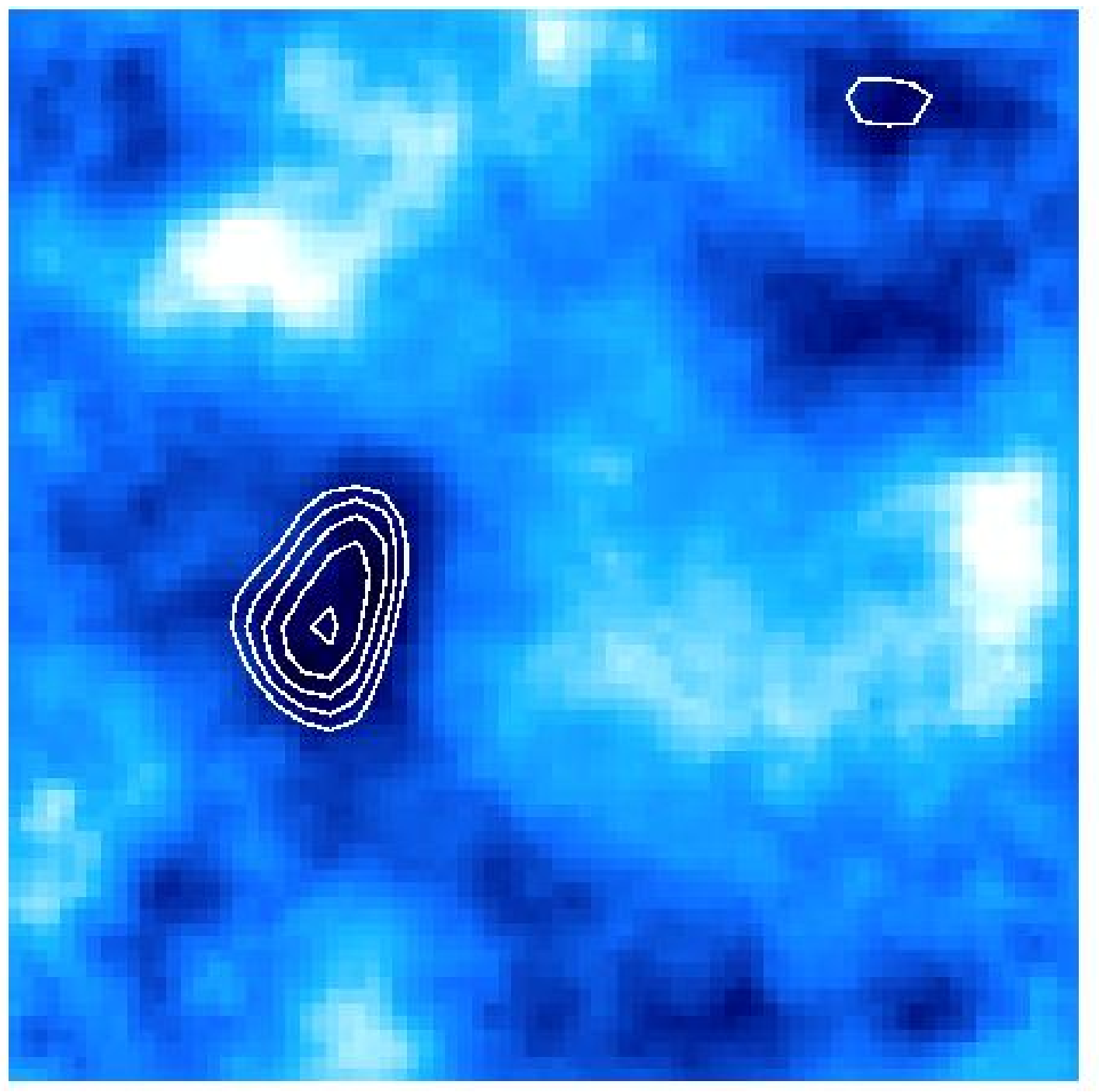}{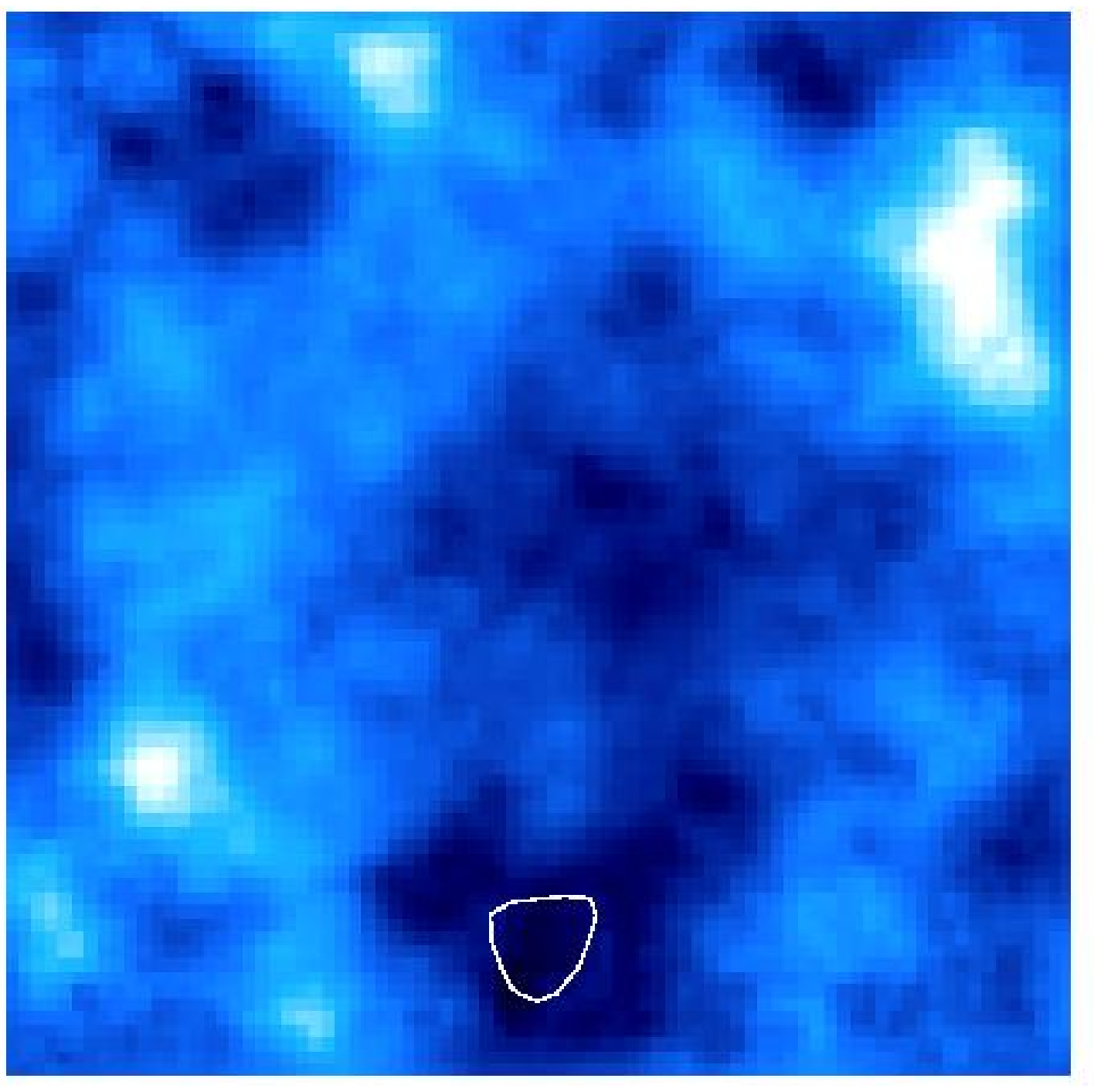}
\caption{Mass (left) and randomized (right) maps for subfield
F4p21. There are approximately $20$K objects in this $35^{\prime}
\times 35^{\prime}$ region, or $\sim 16$ per square arc min.  North is
Up, East is to the Left.}
\label{fig:cl1055maps}
\end{figure}

\begin{figure}[h]
\centering \plottwo{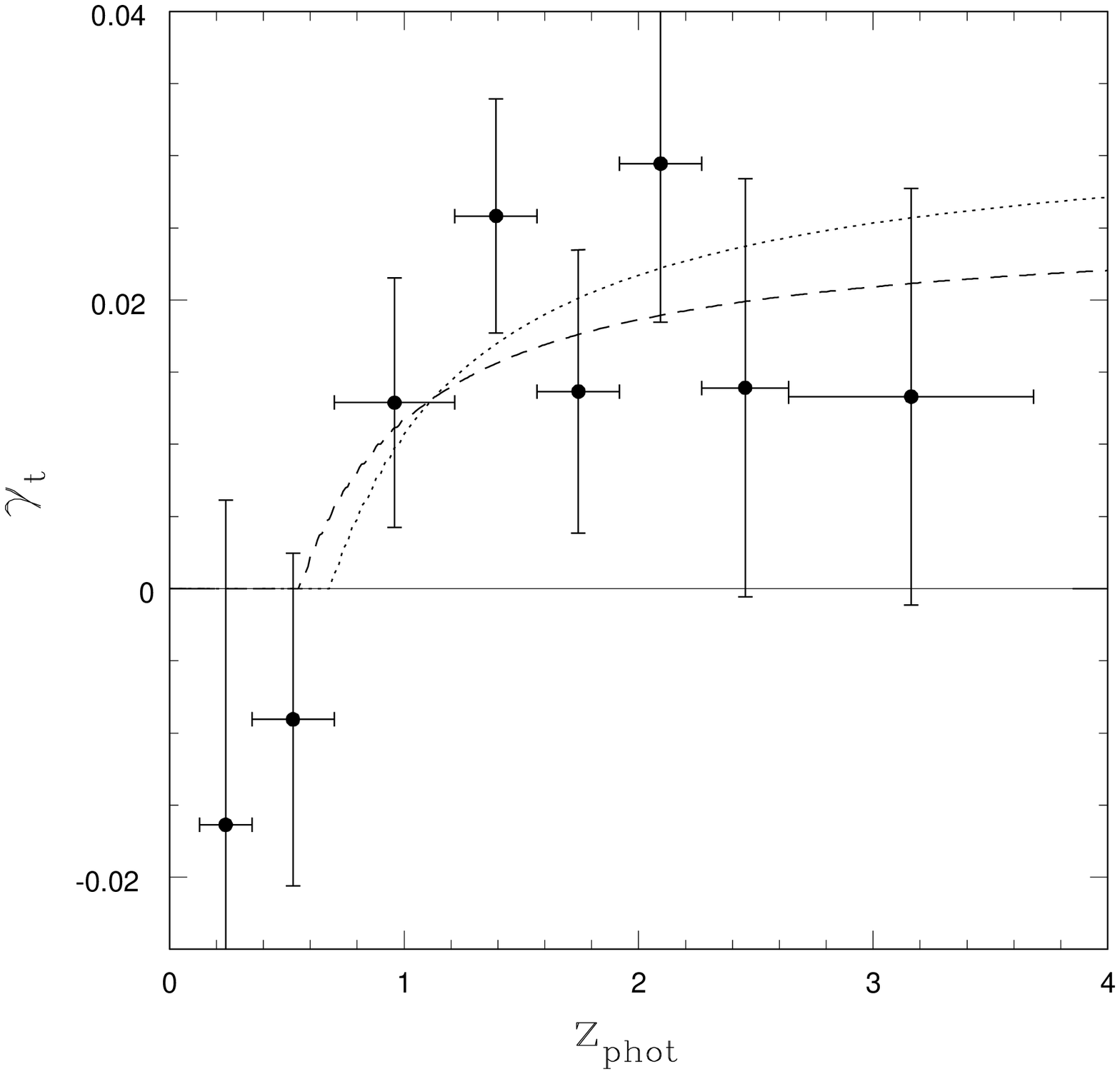}{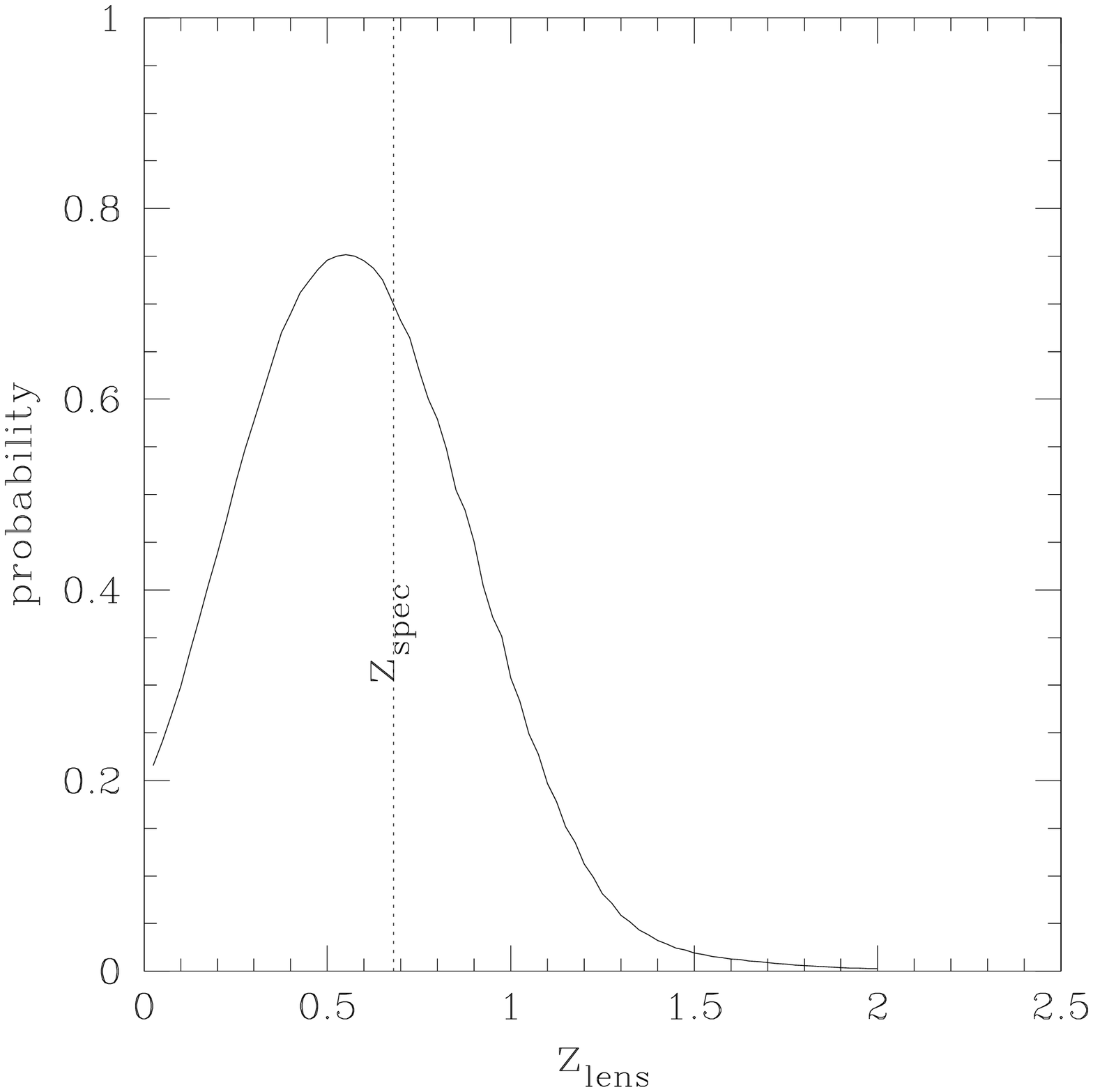}
\caption{Left: Tangential shear, around the mass peak, as a function of source photometric redshift. The dotted curve shows the best-fit lens fixed at the spectroscopic redshift of 0.68, and the dashed curve shows the best fit when the lens redshift is allowed to vary ($z=0.55$). Right: Lens redshift probability distribution. The peak is at $z=0.55$ and the rms is 0.15.}
\label{fig:cl1055shear}
\end{figure}

\begin{figure}[h]
\plotone{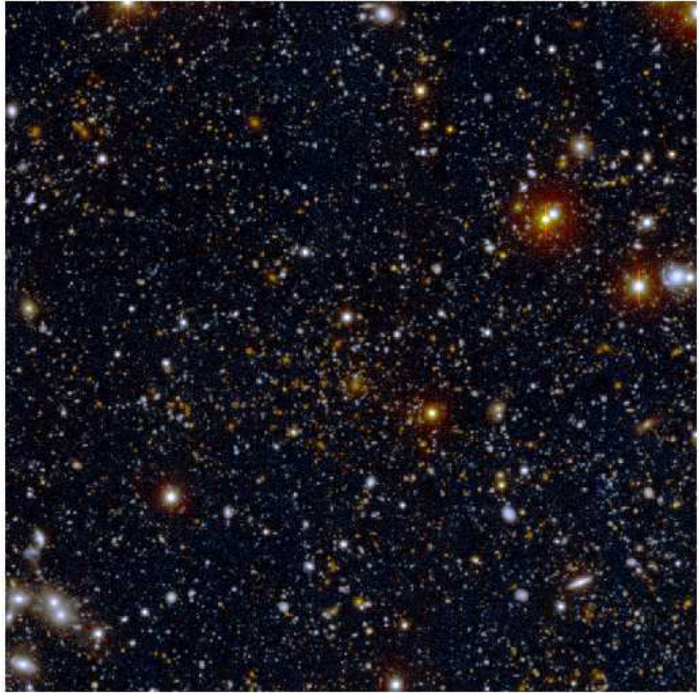}
\caption{A 4$^{\prime}$ square section of the $R$ image, centered on the BCG.  The BCG is $R=20.6$, and the faintest galaxies visible in this reproduction are $R \sim 26$.  North is up, and east to the left.}
\label{fig:cl1055image}
\end{figure}

The first shear-selected cluster with tomography analysis was
presented in a paper by Wittman \etal~2001 in a pilot project for the
DLS.  The DLS, covering $\sim$ 50 times more area, should yield a
significant sample.  In Wittman \etal~2003 the DLS team presents the
first shear-selected cluster from the DLS data, with tomography
analysis.  Figure \ref{fig:cl1055maps} shows the detection of the
cluster in a mass map. The tomography is shown in Figure
\ref{fig:cl1055shear}: the left panel presents the mean tangential
shear for independent redshift bins; and the right panel indicates the
lens (cluster) redshift probability distribution. The best-fit lens
redshift is $0.75 \pm 0.46$, and spectroscopic observations have
confirmed it to be a massive ($\sigma_v = 980$ km s$^{-1}$) galaxy
cluster at $z=0.68$.  Figure \ref{fig:cl1055image} shows a $4^\prime
\times 4^\prime$ image of the cluster.

\eject

\section{Future Work}

\begin{figure}
\centering
\includegraphics[width=.45\columnwidth]{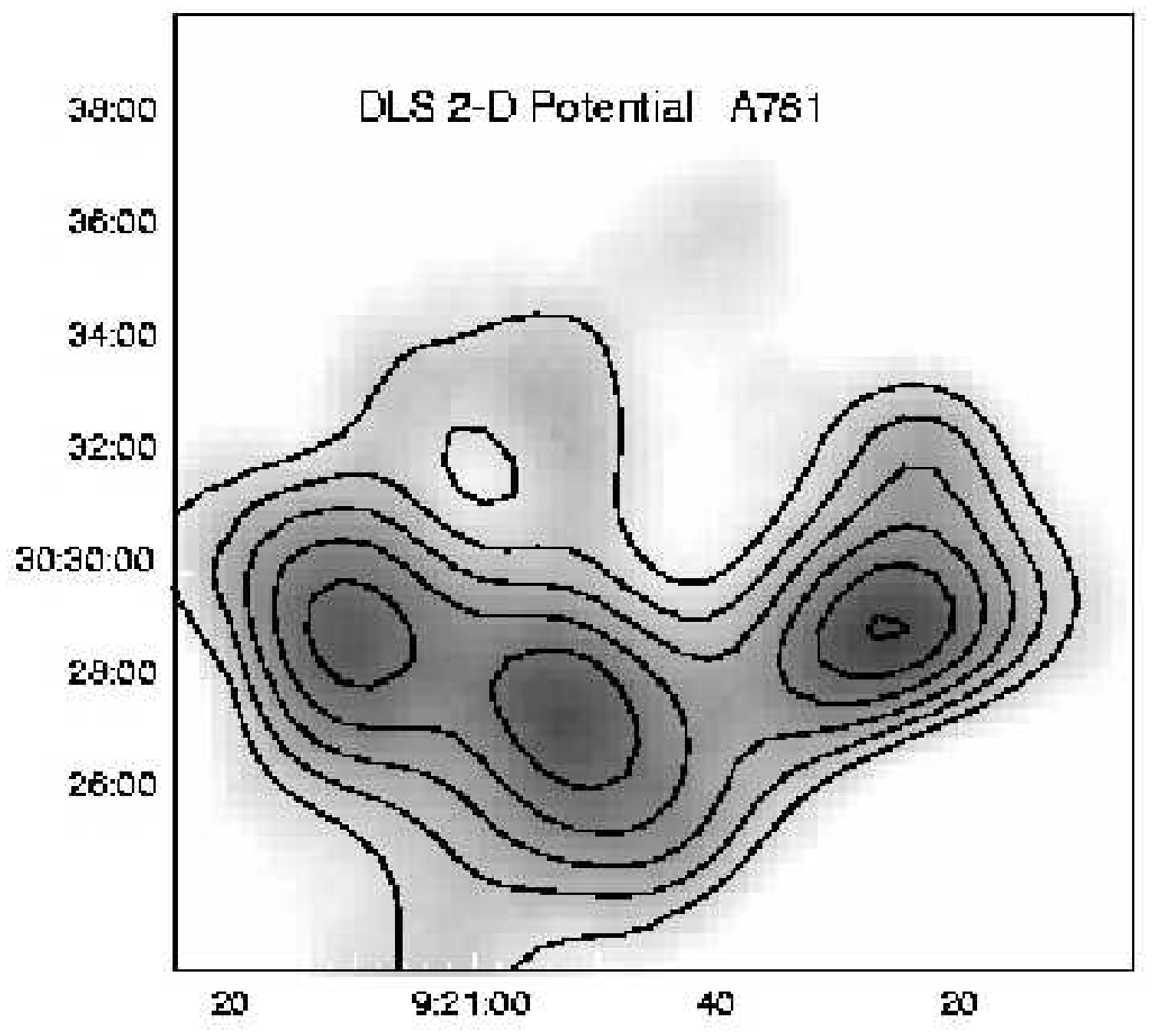}
\includegraphics[width=.40\columnwidth]{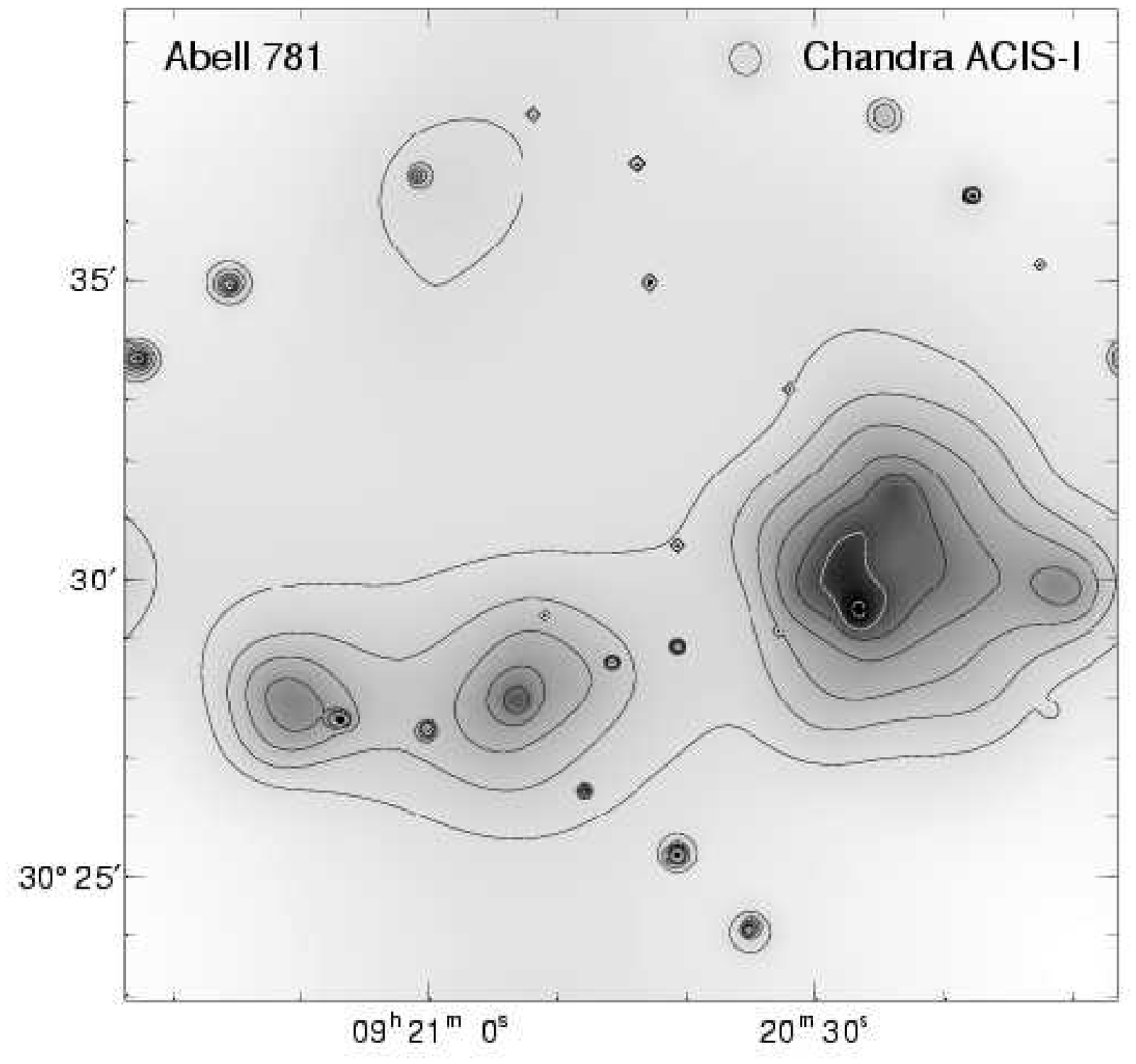}
\caption{Maps of the gravitational potential (left) and \chandra~X-ray
emission (right) for Abell 781. Note the difference in the
distributions shown in the two maps. While the X-ray emission is
predominantly concentrated in the rightmost peak, the potential map
shows that a significant fraction of the mass is contained in the
lefthand structures.}
\label{fig:a781}
\end{figure}

A preliminary sample of 8 clusters detected in the DLS is being
observed with \chandra~and 4 more are scheduled to be observed with \XMM~
(PI. Prof. John P. Hughes). The X-ray data will allow us to quantify
the degree to which baryonic matter traces the dark matter
distribution. The DLS provides a unique sample of clusters whose
selection is unaffected by the distribution of baryons in the cluster,
and with this sample we are finding provocative indications that the
mass and gas distributions exhibit pronounced differences. Figure
\ref{fig:a781} shows a DLS potential map for Abell 781, and the X-ray
emission in the same region. Clear evidence for differences in the
spatial distributions of baryonic and non-baryonic content are
observed.

We have also applied for HST time to obtain higher resolution mass
maps for a few of these clusters (PI Dr. Anthony Gonzales). The ACS
observations will allow us to construct higher resolution
mass/potential maps for this clusters ($\sim$50 $h^{-1}$ kpc compared
to 400 $h^{-1}$ kpc resolution for the DLS data). With mass maps of
such resolution, we will measure the cross-correlation of mass and
light on scales of 0.05-1.5 $h^{-1}$ Mpc and will also quantify the
radial dependence of the bias, which is of critical importance for
derivations of the total cluster baryon fraction.

\section{Conclusions}

The DLS team has already shown that shear-selection is effective at
finding galaxy clusters (Wittman \etal~2001, Wittman \etal~2003).  In
addition to detecting the shear, the multiband imaging produces
photometric redshifts for the clusters.  Indeed, combining shear
measures with photometric redshift estimates {\it for the background
galaxies}, one can obtain a redshift estimate for the lens {\it
independently} of the luminous output of the cluster. This opens the
way for construction of a completely baryon-independent cluster sample
from the DLS.

\section{Acknowledgements}

We thank the KPNO and CTIO staffs for their invaluable assistance.
NOAO is operated by the Association of Universities for Research in
Astronomy (AURA), Inc. under cooperative agreement with the National
Science Foundation.

\end{document}